\documentclass[12pt]{article}
\usepackage{graphicx}

\begin{document}

 \title{The Effect of Partial Coherence \\on the Statistics of Single-Photon
Pulses\\ Propagating in the Atmosphere}
\author{G.P. Berman $^{1}$\footnote{Corresponding
author: gpb@lanl.gov} ~and A.A. Chumak$^{1,2}$
\\[3mm]$^1$
 Los Alamos National Laboratory, Theoretical Division,\\ Los Alamos,
NM 87545
\\[5mm] $^2$  Institute of Physics of
the National Academy of Sciences\\ Pr. Nauki 46, Kiev-28, MSP 03028
Ukraine
\\[3mm]\rule{0cm}{1pt}}
\maketitle \markright{right_head}{LAUR-06-6989}

\begin{abstract}

The photon density operator function is used to describe the propagation
of single-photon pulses through a turbulent atmosphere. The effects of
statistical properties of photon source and the effects of a
random phase screen on the variance of photon counting are studied.
A procedure for  reducing the total noise is discussed. The physical
mechanisms responsible for this  reduction are explained.

\end{abstract}







\section{Introduction}

Fluctuations of the atmospheric refractive index caused by turbulent
eddies considerably limit the performance of long-distance optical
communication systems. An initially coherent laser beam acquires
some properties of Gaussian statistics in the case of a long 
propagation path or strong turbulence. The tendency of the
scintillation index (defined as a normalized intensity variance) to
approach asymptotically the level of unity in this case is a
distinct manifestation of gaussianity of the field statistics. In the
literature this effect, known first from experimental studies of
Gracheva and Gurvich \cite{grach}, is referred to as the saturation of
fluctuations. (See, for example, \cite{tatar}-\cite{kravt}.) In that
way, the strong influence of the refractive index fluctuations on the
radiation field statistics is emphasized.

It has been shown quite generally \cite{plon}-\cite{lee} that the
normalized intensity variance approaches unity for any source
distribution as the turbulence strength tends to infinity. This
result is valid for any degree of coherence of the source, provided
the response time of the recording instrument is short compared with the
source coherence time (fast detector). In the opposite case of a slow
detector, the scintillation index can decrease for partially
coherent beams. The effect is not very strong for  temporal
partial coherence \cite{fant80}, \cite{fan77}-\cite{plon79}, but is
very pronounced in the case of spatial partial coherence. (See papers 
\cite{bana54}-\cite{ber}, which deal with stationary
beams.)

The purpose of the present paper is to study the effects of partial
spatial coherence on the statistics of the detected photons when photons
are generated as individual pulses of electromagnetic field
propagating in the earth's atmosphere. Moreover, the effects of the
initial statistics of photons on fluctuations of the detector counts
will be elucidated.

The case of single-photon pulses is of special interest for quantum
cryptography because individual photons are carriers of information
bits in several basic strategies for free-space quantum key
distribution. (Practical free-space quantum key distribution is
described in references \cite{but}-\cite{mar}.) Fluctuations of the
detector counts for the case of single-photon pulses were studied both
theoretically and experimentally in \cite{mil}. In contrast to
\cite{mil} where two limiting cases (plane-wave and spherical-wave
approximations) were studied, our consideration takes into account
the actual beam diameter. Also we will not restrict ourselves with
the case of small fluctuations of the radiation field as in
\cite{mil}. The case of strong fluctuations will be studied as
well. The method of photon distribution function in phase space
developed in \cite{ber} is somewhat generalized here to apply to 
the case of laser pulses.

\section{The photon distribution function, and pulse propagation in the atmosphere}

Similar to  \cite{ber}, we proceed from the quantum version of
Hamiltonian, $H$, of photons in a medium with a fluctuating refractive
index, $n({\bf r})$ [$n({\bf r})-1<<1$]
\begin{equation}\label{one}
H=\sum_{\bf k}\hbar \omega _{\bf k}b^+_{\bf k}b_{\bf k}-
\sum_{{\bf k},{\bf k^\prime}}\hbar \omega _{\bf k}n_{\bf k^\prime}b^+_
{\bf k}b_{{\bf k}+{\bf k^\prime}}.
\end{equation}
Here the terms describing the zero-point
electromagnetic energy are omitted; the two terms
on the right-hand side describe photons in a vacuum and
the effect of refractive index fluctuations, respectively;  $b^+_{\bf k}$
and $b_{\bf k}$ are creation and annihilation operators of photons with
momentum ${\bf k}$, $\hbar \omega _{\bf k}\equiv \hbar ck$ is the photon
energy; $c$ is the speed of light in a vacuum, and $n_{\bf k}$ is the
Fourier transform of $n({\bf r})$ defined by
\begin{equation}\label{two}
n_{\bf k}=\frac 1V\int dVe^{i{\bf kr}}n({\bf r}),
\end{equation}
where $V\equiv L_xL_yL_z$ is the normalizing volume.

Eq. (\ref{one}) is valid in the limit of small wave-vectors ${\bf
{k^{\prime}}}$ ($k^{\prime}\ll k$). This means that the scale of
spatial inhomogeneity of turbulence is much greater than the
wavelength of the radiation. Also we assume here that the initial
polarization of light remains unaffected by the turbulence throughout
the distance of propagation. The depolarization of light due to
atmosphere turbulence is very small. (See, for example, \cite{stroh}
and \cite{coll} where this effect was studied.)

The photon distribution function is defined by
\begin{equation}\label{three}
f({\bf r},{\bf q},t)=\frac 1V\sum_{\bf k}e^{-i{\bf kr}}b^+_{{\bf q}+
{\bf k}/2}b_{{\bf q}-{\bf k}/2}.
\end{equation}

The operator function $f({\bf r},{\bf q},t)$ is the photon density
in six-dimensional (${\bf r},{\bf q}$) phase space. We will
use it to describe light pulses with characteristic sizes, $l$, much
greater than the wavelength of the radiation, $\lambda $. In this
case, it is convenient to restrict the sum over ${\bf k}$ by some $k_0$
($k<k_0<<q_0$, $k_0>2\pi /l$, where $q_0$ is the wave vector
corresponding to the central frequency $\omega _0$ of radiation,
$\omega _0=cq_0$). Then, the kinetic equation for the distribution
function is governed by
\begin{equation}\label{fo}
\{ \partial_t +{\bf c_q}\partial_{\bf r}+{\bf F}({\bf r})\partial_{\bf q}\}
f({\bf r},{\bf q},t)=0,
\end{equation}
where ${\bf F}({\bf r})=
\omega _0\partial_{\bf r}n({\bf r})$ and
${\bf c_q}=\partial \omega _q/\partial {{\bf q}}$.
 When deriving Eq. (\ref{fo}) we have considered the
refractive index to be a slowly varying function of the coordinate,
${\bf r}$. (See more details in \cite{ber}.) In this case the effect
of the turbulence on $f$ is represented by the random force ${\bf
F}({\bf r})$, that can be seen in Eq. (\ref{fo}).

The general solution of Eq. (\ref{fo}) is 
\begin{equation}\label{fi}
f({\bf r},{\bf q},t)=\phi \Bigg\{{\bf r}-\int _{t_0}^tdt^\prime\frac
{\partial
{\bf r}(t^\prime)}{\partial t^\prime};{\bf q}-\int_{t_0}^tdt^\prime\frac
{\partial{\bf q}(t^{\prime})}{\partial t^\prime};t_0\Bigg\},
\end{equation}
where the functions ${\bf r}(t^\prime )$ and ${\bf q}(t^\prime)$ are
photon ``trajectories" defined by the equations of motion
\[ \frac {\partial {\bf r}(t)}{\partial t}={\bf c}[{\bf q}(t)], \]
\begin{equation}\label{si}
\frac {\partial {\bf q}(t)}{\partial t}={\bf F}[{\bf r}(t)],
\end{equation}
and the corresponding initial conditions. The last equations follow from the
requirement that the trajectories in Eq. (\ref{fi}) ${\bf r}(t^\prime )$
and ${\bf q}(t^\prime)$ pass through the point ${\bf r},{\bf q}$ at
 $t^\prime =t$
[i.e. ${\bf r}(t^\prime =t)={\bf r}, {\bf q}(t^\prime =t)={\bf
q})$].  The initial value of $f({\bf r},{\bf q},t)$, is $\phi ({\bf
r},{\bf q},t_0)$
\begin{equation}\label{se}
\phi ({\bf r},{\bf q},t_0)=f({\bf r},{\bf q},t_0)=\frac 1V\sum_{\bf k}
e^{-i{\bf kr}}(b^+_{{\bf q}+{\bf k}/2}b_{{\bf q}-{\bf k}/2})|_{t=t_0}
\equiv \sum_{\bf k}e^{-i{\bf kr}}\phi ({\bf k},{\bf q},t_0).
\end{equation}

It is convenient to set $t_0$ equal to the instant just after photon
exits the source (as shown in Fig. 1). These photons have not been
affected by the atmospheric turbulence and their statistics is
determined by the source properties only. The operators, $b_{\bf q}^+$, and
$b_{\bf q}$, describe amplitudes of the field, $E^{atm}$, outgoing from the source.
The correspondence between free-space and generated modes
can be established using the following reasoning. The field in the
atmosphere is given by
 \begin{equation}\label{sev1}
E^{atm}({\bf r})=i\sum_{\bf q}\Bigg(\frac {2\pi \hbar \omega _{\bf
q}}{V}\Bigg)^{1/2} [e^{i{\bf qr}}b_{\bf q}-e^{-i{\bf qr}}b^+_{\bf q}].
\end{equation}
On the other hand, the outgoing field localized in the vicinity of the laser aperture can be expressed in terms of the
\begin{figure}[ht]
\centering
\includegraphics{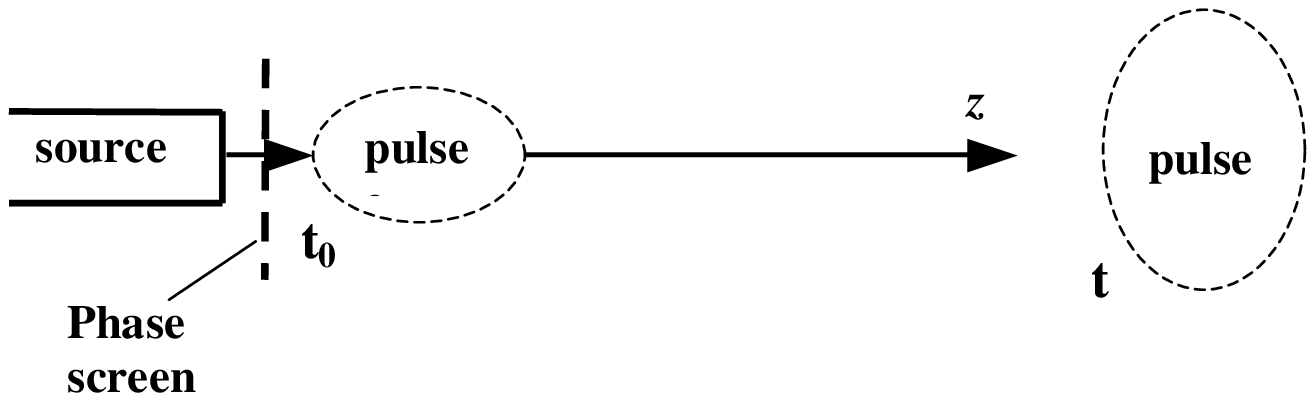}
\caption{The scheme of the communication channel.}
\end{figure}
 laser mode  creation
and annihilation operators, $b^+$ and $b$, and the normalized functions, $\Phi
^*({\bf r})$ and   $\Phi ({\bf r})$, describing the spatial distribution of the
field, as in Ref. \cite{mil07}:
\begin{equation}\label{seve}
E^{s}({\bf r})=i(2\pi \hbar\omega _0)^{1/2} [ \exp({iq
_{0}z})\Phi({\bf r})b-\exp({-iq _{0}z})\Phi ^*({\bf r})b^+].
\end{equation}
The function, $\Phi({\bf r})$, can be chosen as
\begin{equation}\label{seven}
\Phi({\bf r})=\bigg (\frac 2\pi \bigg )^{3/4}(r_0^2r_z
)^{-1/2}\exp\bigg (-\frac {r^2_\perp }{r_0^2}-\frac
{(z-z_0)^2}{r_z^2}\bigg ),
\end{equation}
where $r_\perp ^2=x^2+y^2;~ z_0$ is the position of the pulse center
at $t=t_0$; and $r_0$ and $r_z$ describe the aperture radius and the
length of light pulse, respectively. Then, from the condition
$E^{atm}({\bf r},t_0) =E^{s}({\bf r},t_0)$ we obtain
\begin{equation}\label{ei}
b_{\bf q}(t_0)=\frac {b(t_0)}{\sqrt V}\int d{\bf r}e^{-i({\bf q}-{\bf q}_0){\bf r}}
\Phi ({\bf r}).
\end{equation}
It should be noted that the condition $E^{atm}=E^{s}$ is not 
rigorous. We ignore here vacuum fields arising from other modes
of the resonator as well as the zero-point free-space waves reflected
from the output window. These fields, providing correct commutation
relations for the operators $b_{\bf q}^+, ~b_{\bf q}$, do not
contribute to the detected number of photons. There are no photons
in these fields. (Direct detection is assumed.) Thus, the
corresponding ``vacuum" terms are omitted in Eq. (\ref{ei}) as
being irrelevant to our problem.

The effects of the phase screen can be included by introducing the
Multiplier, $e^{i\varphi ({\bf r}_\perp )}$, into the integrand of Eq.
(\ref {ei}), where $\varphi ({\bf r}_\perp )= {\bf ar}_\perp $ and
$\bf a$ is a Gaussian random variable with a covariance
$\langle(a_{x,y})^2\rangle=2\lambda ^{-2}_c$. This results in the
following relation for the average of the exponent $\langle e^{i{\bf
a}{\bf r}_\perp }\rangle=e^{-r_\perp ^2 \lambda _c^2}$. This relation will 
be useful in our further analysis. Also, the fluctuations of the
refractive index are usually considered as Gaussian variables with a
spatial Fourier-component of the correlation function $\langle
n({\bf r})n({\bf r}^\prime )\rangle_{\bf g}\equiv \psi ({\bf g})$
given by
 \begin{equation}\label{ni}
\psi ({\bf g})=0.033C_n^2\frac {\exp[-(gl_0/2\pi
)^2]}{[g^2+L_0^{-2}]^{11/6}}.
\end{equation}
Eq. (\ref{ni}) is refereed to as the von Karman spectrum. $L_0$ and
$l_0$ are the outer and inner scales sizes of the turbulent eddies,
respectively. In atmospheric turbulence, $L_0$ can range from 1 to
100 meters, and $l_0$ is of the order of several millimeters.
$C_n^2$ is the index-of-refraction structure constant. In
most physically important cases the quantity $L_0^{-2}$ in the
denominator of Eq. (\ref{ni}) can be omitted. In this case, the von
Karman spectrum is reduced to the Tatarskii spectrum \cite{tatar}.

Eqs. (\ref{fi}) and (\ref{se}, \ref{seven}-\ref{ni}) are sufficient
to determine the beam intensity
\begin{equation}\label{te}
\langle I({\bf r},t)\rangle=c\sum_{\bf q}\hbar \omega _{\bf
q}\langle f({\bf r},{\bf q},t)\rangle
\end{equation}
at any ${\bf r}$ and $t$. Represent Eq (\ref{fi}) in the form

\[f({\bf r},{\bf q},t)=\sum_{\bf k}\exp\Bigg\{-i{\bf k}\{ {\bf r}-{\bf c_q}(t-t_0)+\frac c{q_0}\int
_{t_0}^tdt^\prime
(t^\prime -t_0){\bf F}_\perp  [{\bf r}(t^\prime )]\} \Bigg\}\times \]
\begin{equation}\label{el}
\varphi _{\bf k}\Bigg\{{\bf q}-\int _{t_0}^tdt^\prime
{\bf F}_\perp  [{\bf r}(t^\prime )];t_0\Bigg\}.
\end{equation}
This form is more convenient for obtaining the explicit form of Eq.
(\ref{te}). We have neglected the changes of the longitudinal photon
momentum ($q_z$) caused by the turbulence, because of their
negligible contribution to the initial momentum ($\approx q_0$) for
almost any reasonable propagation path.

The photon ``trajectories" ${\bf r}(t^\prime )$ in the arguments of
${\bf F}_\perp$ can be approximated by straight lines, ${\bf
r}(t^\prime )={\bf r}+ {\bf c_q}(t^\prime -t)$. Thus, we disregard
the variation of photon momentum at distances of the order of the
turbulence correlation length, $L_0$, which is assumed to be much less
than the total propagation path, $c(t-t_0)$. Then, the explicit
expression for $\langle I\rangle$ is given by
\begin{equation}\label{tv}
\langle I({\bf r},t)\rangle=\sqrt {\frac 2\pi}\frac {c\hbar \omega
_0}{\pi R^2r_z}\exp\Bigg\{-\frac {r_\perp ^2}{R^2} -2\frac
{z_{eff}^2}{r_z^2}\Bigg\},
\end{equation}
where \[z_{eff}=z-z_0-c(t-t_0), ~ R^2=\frac {r_0^2}{2}\Bigg
\{1+\Bigg[ \frac {2c(t-t_0)}{q_0r_0r_1}\Bigg]^2+ \frac
{8c^3(t-t_0)^3T}{r_0^2}\Bigg \},\] \[r_1^2=\frac
{r_0^2}{1+2r_0^2\lambda _c^{-2}},\] and
\[T=0.558C_n^2l_0^{-1/3}.\] $z_{eff}$ and $R$ are the distance to
the pulse center at time $t$ and the beam radius,
respectively.

The effects of partial coherence on the beam radius is represented by
the quantity, $r_1$, which enters the second term in expression for
$R^2$. This term describes the diffraction broadening of the beam in
the course of its propagation. The third term is due to turbulence.
It dominates at large distances. It does not depend on the
initial partial coherence. Therefore, the beam radius is not
sensitive to the presence of phase screen when $c(t-t_0)\rightarrow
\infty$. In this connection an important question for our analysis
arises: Is there any effect of the partial coherence on the
photon-counts statistics at large propagation paths? This issue is
considered in the next section.

Eq. (\ref{tv}) is derived for a single-photon pulse, i.e. for
$\langle b^+b\rangle=1$. For arbitrary photon number per pulse,
$N_{pulse}$, the coefficient in front of the exponent in Eq. (\ref{tv})
should be multiplied by $N_{pulse}$. The  case of a homogeneous beam
(see, for example, \cite{ber}) can be considered by assuming
$r_z\rightarrow \infty$ and by renormalizing the coefficient in front of
the exponent. One can easily see that the beam radius is increased
with the distance (or propagation time $(t-t_0)$) similar to the
case of the stationary intensity. In contrast, the pulse length
remains unaffected by the turbulence, within the formalism described
above.

\section{Fluctuations of photon counting}

As before, we assume that the background radiation noise is
negligible. Also, the detector area is taken to be small compared
with the beam width. This is quite reasonable for systems with long
transmission distances. The counting interval, $T_p$, is much greater
than the pulse duration $\tau _p$ ($\tau _p\sim r_z/c,~\\T_p>>\tau
_p$), as shown in Fig. 2. (In experiments \cite{mil}, the
radiation was in the form of $1 ns$ laser pulses with a repetition
rate $1$ MHz and the interval $T_p=10 ms$.)
\begin{figure}[ht]
\centering
\includegraphics{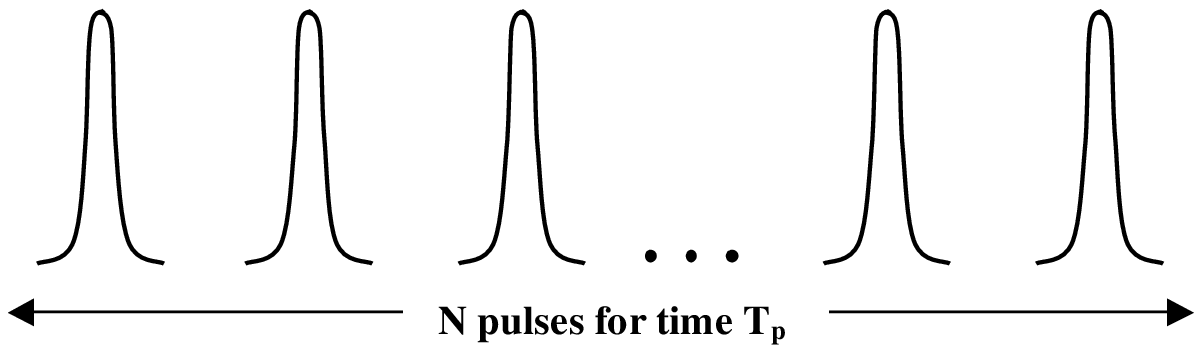}
\caption{A sequence of single-photon pulses.}
\end{figure}

We start with the definition of the mean-square of the photon
counts, $n$, for the time interval, $T_p$. Usually, this is represented
as a sum of two terms
\begin{equation}\label{thi}
\langle n^2\rangle=\langle n\rangle+\langle n(n-1)\rangle,
\end{equation}
where $\langle n\rangle=\alpha N$, and $\alpha $ describes the light
collection and the quantum detection efficiency. It is the ratio of
the numbers of detected $\langle n\rangle$ and generated photons,
$N$. The average number $\langle n\rangle$ is determined by the
integral of light intensity $I({\bf r},t)$ as
\begin{equation}\label{four}
\langle n\rangle=\eta \int _{t-T/2}^{t+T/2}dt^\prime \langle I({\bf
r},t^\prime )\rangle,
\end{equation}
where $\eta $ describes the detector efficiency. By substituting Eq.
(\ref{tv}) into Eq. (\ref{four}), and assuming  the detector is
at the beam center, ${\bf r}_\perp =0$,  we have
\begin{equation}\label{fourt}
\langle n\rangle=\eta N\frac {\hbar \omega _0}{\pi R^2}.
\end{equation}
Thus, we see that $\alpha $ and $\eta$ are related by 
\[\alpha =\eta \frac {\hbar\omega _0}{\pi R^2}.\]

The second term in Eq. (\ref{thi}) is
determined by \cite{man}
\begin{equation}\label{fif}
\langle n(n-1)\rangle=\Big <:\Big \{\eta \int
_{t-T_p/2}^{t+T_p/2}dt^\prime I({\bf r},t^\prime ) \Big \} ^2:\Big
>,
\end{equation}
where the symbol $\langle:...:\rangle$ indicates normal ordering of
the operators $b^+_{\bf q}(t^\prime)$ and $b_{{\bf q}^\prime }(t^{\prime
\prime})$ with subsequent averaging. In our case these operators
enter the definition of the intensity $I({\bf r},t^\prime )$ and,
hence, the right part of Eq. (\ref{fif}).

Each integration over $t^\prime$ within the total interval $T_p$ is
reduced to a sum of $N$ independent integrations within much
smaller intervals $\Delta t_i$ [$(r_z/c)<<\Delta
t_i<<T_p/N,i=1,2,..N$], which correspond to instants when the pulses
cross the detector plane. Then Eq. (\ref{fif}) is reduced to
\[\langle n(n-1)\rangle=\eta ^2N(N-1)\Bigg (\frac {\hbar \omega _0c}V\Bigg )^2\sum_{{\bf q},{\bf k}}
\sum_{{\bf q}^\prime ,{\bf k}^\prime}\int dt^\prime\int dt^{\prime \prime}
e^{-i({\bf k}+{\bf k}^\prime ){\bf r}}\times \]
\begin{equation}\label{six}
\big < b^+_{{\bf q}+{\bf k}/2}(t^\prime) b^+_{{\bf q}^\prime+{\bf k}^\prime /2}(t^{\prime \prime})
b_{{\bf q}^\prime -{\bf k}^\prime /2}(t^{\prime \prime})b_{{\bf q}-{\bf k}/2}(t^\prime)\big>,
\end{equation}
where integrations are within any two different intervals $\Delta
t_i$. The coefficient $N(N-1)$ indicates the number of such
intervals. Coinciding intervals do not contribute to Eq. (\ref{fif})
because of the zero value of the intensity-intensity correlation for a
single-photon pulse.

Not all terms in the sum over ${\bf q},~{\bf k},~{\bf
q}^\prime,~{\bf k}^\prime$ contribute significantly to $\langle
n(n-1)\rangle$. The analysis in \cite{ber} shows that at large but
finite propagation distances the terms with small values of (i)
$k,~k^\prime$ or (ii) $|{\bf q}^\prime -{\bf q}+({\bf k}+{\bf
k}^\prime )/2|,~|{\bf q} - {\bf q}^\prime+({\bf k}+{\bf k}^\prime
)/2|$ (almost diagonal terms) are the most important.
 The evolution of
such terms can be described in a manner similar to the case of the evolution of $f$. Then,
the explicit form of the expression (\ref{six}) is given by
\[\langle n(n-1)\rangle=\eta ^2N(N-1)(2\pi )^5\Big (\frac {\hbar \omega _0r_1^2r_z}{V^2}\Big )^2
\sum_{{\bf q},{\bf k}}\sum_{{\bf q}^\prime ,{\bf k}^\prime}
e^{-\big [\Delta q_z^2+\Delta q_z^{\prime 2}+(k_z^2+k_z^{\prime 2})/4\big ]r_z^2/2}\times\]
\[ \Big <\Big [\delta (k_z)\delta (k_z^\prime )e^{-(Q^2+Q^{\prime 2})r_1^2/2-(k_\perp ^2+
k_\perp ^{\prime 2})r_0^2/8}+\]
\[\delta (q_z-q_z^\prime)\delta (k_z+k_z^\prime )e^{-\big[({\bf Q}+{\bf Q}^\prime)^2+({\bf k}_\perp -
{\bf k}^\prime _\perp )^2/4\big ]r_1^2/4-\big [({\bf Q}-{\bf Q}^\prime)^2+({\bf k}_\perp +
{\bf k}^\prime _\perp )^2/4\big ]r_0^2/4}\Big ]\times \]
\begin{equation}\label{sev}
e^{-i\big \{{\bf k}_\perp \big [{\bf r}-{\bf c}({\bf q})t_z+\frac c{q_0}\int _0^{t_z}dt_1t_1{\bf F}
\big ({\bf r}_{\bf q}(t_1)\big )\big ]+{\bf k}^\prime _\perp \big [{\bf r}-{\bf c}({\bf q}^\prime )t_z+
\frac c{q_0}\int _0^{t_z}dt_1t_1{\bf F}\big ({\bf r}_{{\bf q}^\prime} (t_1)\big )\big ]\big \}}\Big >,
\end{equation}
where $\Delta q_z=q_z-q_o;~ {\bf Q}={\bf q}_\perp -\int _0^{t_z}dt_1
{\bf F}_\perp \big ({\bf r}_{\bf q}(t_1)\big )$; ~${\bf Q}^\prime
={\bf q}_\perp ^\prime -\int _0^{t_z}dt_1 {\bf F}_\perp \big ({\bf
r}_{{\bf q} ^\prime }(t_1 )\big )$, and the function ${\bf r}_{\bf
q}(t_1)$ is the particle trajectory which passes through the point (${\bf
r}$, ${\bf q}$) at the instant $t_1 =t_z\equiv t-t_0$.

Summation over the variables $q_z,~q_{z}^\prime,~k_z,~k_{z}^\prime $ can
be easily performed. After that, the remaining sum coincides exactly with
the corresponding sum in the right part of Eq. (33) of reference
\cite{ber}. Hence, we have
\begin{equation}\label{eight}
\langle n(n-1)\rangle=\alpha ^2N(N-1)(1+\sigma ^2),
\end{equation}
where $\sigma ^2$ is the scintillation index for the stationary beams calculated in \cite{ber}
using the formalism of photon distribution function that is similar to the presented here.
It follows from Eqs. (\ref{six}) and (\ref{eight}) that
the normalized variance of photon counting can be written in the form
\begin{equation}\label{nint}
\frac {\langle n^2\rangle-\langle n\rangle^2}{\langle
n\rangle^2}=\frac {1-\alpha}{\langle n\rangle}+\sigma
^2\Bigg(1-{1\over N}\Bigg).
\end{equation}
Eq. (\ref{nint}) was obtained for the case of long propagation
distances. Nevertheless, it has the same form in the opposite limit
of short distances (or weak turbulence) when perturbation
methods like Rytov's approach \cite{tatar},~\cite{ber} are
applicable.

The first term in the right part of Eq. (\ref{nint}) is due to the
discrete (quantum) nature of the photon field. When $\alpha
\rightarrow 0$, it is reduced to $1/\langle n\rangle$ which is known
in the literature as the shot-noise limit (or the standard quantum
limit). The coefficient ($1-\alpha $) in the numerator arises from the
$\langle n(n-1)\rangle$ term in Eq. (\ref{thi}) and is evidence
of the nonclassicality of the light. The presence of the factor
($1-\alpha $) in Eq. (\ref{nint}) is typical for squeezed
photon-number radiation. In our case, the total photon number, $N$, is
considered to be a constant. For photocount statistics, this
situation is equivalent to the case of photon number state (Fock
state) of the light. In the hypothetical case of $\alpha =1$, the
quantum term vanishes. This case corresponds to the physical situation in which  all generated photons reach the detector aperture and each
photon produces a photocount. Of course, this situation is
impossible for long propagation distances because the beam radius
becomes much greater than the receiver aperture. In addition, the
detection quantum efficiency is always less than $100\% $.

 The second term in Eq. (\ref{nint}) can be
interpreted as being caused by atmospheric turbulence. The scintillation index determines
its relative contribution to the total noise. In the general case, $\sigma ^2$ depends
on the propagation distance, the radius of the source aperture, turbulence strength, and the
correlation length, $\lambda _c$.
\begin{figure}[ht]
\centering
\includegraphics{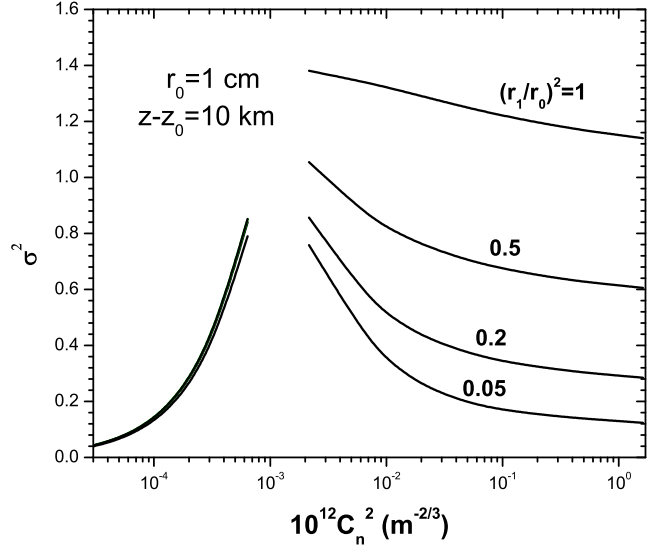}
\caption{Dependence of scintillation index on the turbulence strength $C_n^2$.}
\end{figure}
Fig. 3 shows that the value of $\sigma ^2$ can be suppressed
considerably by decreasing the initial coherence length,
$\lambda _c$. Significant effect is achieved even for a moderate
decrease of the dimensionless parameter $(r_1/r_0)^2$. For example,
we see in Fig. 3 an
 almost $50\% $ decrease in $\sigma ^2$ when $(r_1/r_0)^2=1/2$.

When deriving Eqs. (\ref{eight},~\ref{nint}), we have assumed that
the total number of photons $N$ in the interval $T_p$ is constant
(no fluctuations of $N$). In general, $N$ is a fluctuating quantity.
For example, much research uses heavily attenuated laser pulses to approximate a single-photon
source 
\cite{but},~\cite{hug},~\cite{mil},~\cite{hug06}. (Usually the photon
number per pulse is less than 1.) In this case, the random variable $N$ obeys Poisson statistics resulting 
in the average value $\langle
N(N-1)\rangle$ equal to $\langle N\rangle^2$. Then, Eq.
(\ref{eight}) becomes
\begin{equation}\label{twen}
\langle n(n-1)\rangle=\alpha ^2\langle N\rangle^2(1+\sigma ^2).
\end{equation}
Instead of Eq. (\ref{nint}) we have
\begin{equation}\label{twon}
\frac {\langle n^2\rangle-\langle n\rangle^2}{\langle
 n\rangle^2}=\frac 1{\langle n\rangle}+\sigma ^2.
\end{equation}
Comparison of Eqs. (\ref{nint}) and (\ref{twon}) shows the increase of total noise in
the latter case that is due to contribution of the generation-rate fluctuations.

\section{Conclusion}

The central point of the present paper is  the possibility of reducing
the photocount fluctuations. It follows from the previous Section
that a single-photon-on-demand source or squeezed
photon-number light provide lower noise levels than a heavily attenuated laser
source. Also, an additional decrease in photocount fluctuations can be
achieved by means of a random phase screen.

It can be seen that the phase screen can decrease
the count noise for long-distance propagation. In
the opposite case a similar effect can be achieved by improving the
noise characteristics of the source.

The question arises, ``What is the physical nature of the second terms
in Eqs. (\ref{nint}) and (\ref{twon}) for the case of single-photon
pulses?" These terms describe intensity-intensity correlations. At
the same time, one can easily see that this correlation is absent
within any given pulse. Also, different pulses are independent
events. In principle, they can be generated even by different
sources. Different photons do not interact one with another.
Nevertheless, $\sigma ^2\not =0$. This paradox can be explained as
follows. Different photons move in a fixed distribution of the
refractive index. (It was tacitely assumed that the turbulence does
not vary during the integration time $T_p$.) They are affected by
the same random force, ${\bf F}$. Therefore, there is a strong
correlation of photon's trajectories. The probabilities of detection
for any photons propagating within small time interval $T_p$ depend
on the same refractive-index configuration. Hence, the detection
events within this interval $T_p$ are correlated. The main purpose of
the phase screen is to destroy this correlation. When the
characteristic time of phase variation introduced by the phase
screen is of the order $\tau_d<<T_p$ and, in addition, $\lambda
_c<<r_0$, the scintillation index is decreased considerably for long propagation paths. Thus, a rapid spatio-temporal
variation of the phase is an effective method for decreasing the
photon-counting noise.

\section{Acknowledgment}
 We thank B.M. Chernobrod, G.D. Doolen, and V.N. Gorshkov for discussions. This work was carried out under the auspices of the
National Nuclear Security Administration of the U.S. Department of
Energy at Los Alamos National Laboratory under Contract No.
DE-AC52-06NA25396.

\newpage \parindent 0 cm \parskip=5mm





\end{document}